\documentclass[12pt]{article}
\usepackage{amsfonts,amsmath,amssymb}
\usepackage[pdftex]{graphicx}
\usepackage{hyperref}
\usepackage{float}

\newcommand{\pa}{\partial}
\newcommand{\nn}{\nonumber}
\newcommand{\Ord}{{\cal O}}

\makeatletter

\@addtoreset{equation}{section}
\makeatother

\def\href#1#2{#2}
\begin{document}

\begin{titlepage}

\begin{center}

\hfill 
\vskip25mm

\textbf{\LARGE {Large Field Excursions from }}\\[3mm] 
\textbf{\LARGE {Dimensional (De)construction}}\\[14mm] 

{\large Kazuyuki Furuuchi, Suvedha Suresh Naik}\\[2mm]
{\large and}\\[2mm] 
{\large Noel Jonathan Jobu}\\[8mm]

{\sl
{Manipal Centre for Natural Sciences}\\
{Centre of Excellence}\\
{Manipal Academy of Higher Education}\\
{Dr.~T.M.A.~Pai Planetarium Building}\\
{Manipal 576 104, Karnataka, India}
}

\vskip8mm 
\end{center}
\begin{abstract}
An inflation model
based on dimensional (de)construction 
of a massive gauge theory is proposed.
The inflaton in this model
is the ``zero-mode'' 
of
a component of the massive gauge field
in the (de)constructed extra dimensions.
The 
inflaton potential 
originates from the
gauge invariant Stueckelberg potential.
At low energy, the field range of
the inflaton 
is enhanced by a factor
$N^{\frac{d}{2}}$
compared to
the field range of the original fields in the model,
where 
$d$ is the number of the (de)constructed extra dimensions
and $N$ 
is the number of the lattice points 
in each (de)constructed dimension.
This enhancement of the field range
is used to achieve a trans-Planckian inflaton field excursion.
The extension of the mechanism
\textit{``excursions through KK modes''} 
to the case of (de)constructed extra dimensions
is also studied.
The burst of particle productions
by this mechanism may have 
observable consequences
in a region of the model parameter space.
\end{abstract}

\end{titlepage}

\tableofcontents

\section{Introduction}\label{sec:intro}

Dimensional (de)construction
\cite{ArkaniHamed:2001ca,Hill:2000mu}
provides 
a description of (latticized)
extra dimensions
purely in terms of $4D$ QFT.
One of its important applications
is a
purely $4$D construction
\cite{ArkaniHamed:2001nc}
of 
gauge-Higgs unification models
\cite{Hosotani:1983xw,Hosotani:1988bm,Davies:1988wt,Hatanaka:1998yp}.
Most of the mechanisms
for protecting 
the mass of the Higgs field 
against radiative corrections
have also been used to
make the inflaton potential sufficiently flat 
to realize slow-roll inflation.
Therefore, it is natural to 
ask whether 
the (de)construction of extra-natural inflation
\cite{ArkaniHamed:2003wu,Kaplan:2003aj},
which 
incorporates 
the same mechanism 
used in gauge-Higgs unification to inflation, 
provides a viable 
model.
However, it was already noticed 
in \cite{ArkaniHamed:2003wu}
that 
large field inflation
was difficult to achieve 
in (de)constructed extra-natural inflation.
The main obstacle was that
the trans-Planckian inflaton excursion requires
the period of the inflaton potential $2\pi f_4$
to be much bigger than the 
(reduced) $4$D Planck scale $M_P$:
$2 \pi f_4 \gg M_P$.
However, in the (de)constructed model
with one (de)constructed extra dimension,
$f_4$ is related to a symmetry breaking scale
$f$ in the model
as $f_4 = f / \sqrt{N}$,
where 
$N$ is the number of the lattice points
in the (de)constructed dimension.
As we do not have a valid description
of the physics near the quantum gravity scale $M_P$,
the model is restricted to have
$f \ll M_P$,
it follows that $f_4 \ll M_P$.

In this article,
we consider the (de)construction
of a massive gauge field theory.
Due to the (de)constructed version
of the Stueckelberg mechanism,
the model can have gauge invariant potential 
for the gauge field in 
the (de)constructed extra dimensions.
The inflaton is 
the ``zero-mode''
of a component of 
the gauge field
in the (de)constructed extra dimensions.
We consider a scenario in which
the potential of the massive gauge field
in the (de)constructed extra dimensions
dominates the inflaton potential,
circumventing the problem mentioned above.
Moreover,
we find that
the period of the inflaton potential
is enhanced by a factor of $N^{\frac{d}{2}}$
compared with the potential for the original fields,
where $d$ is the number of the (de)constructed extra dimensions
and $N$ is the number of the lattice points
in each (de)constructed dimension.
The enhancement of the field range 
by a large number of fields
in this model
may be reminiscent of N-flation \cite{Dimopoulos:2005ac}.
However, the enhancement mechanisms
are quite different in detail.
In our model, the inflaton 
is a linear combination of the original fields,
and the enhancement 
is due to the canonical normalization of many fields
which have the same functional form of the potential
due to the discrete translational symmetry on the lattice.
On the other hand,
in N-flation model,
the inflaton was essentially the distance in the field space,
and the enhancement of the field range 
is due to the Pythagoras theorem.
In addition, in our model 
the (de)constructed version of KK reduction 
naturally selects the low energy degrees of freedom.

We also extend the rapid particle production mechanism of
\textit{``excursions through KK modes''} 
\cite{Furuuchi:2015foh}
which periodically occurs as the inflaton travels 
a certain field distance,
to the case in which the ``KK'' modes are 
the lattice counterpart of those in continuous extra dimensions.


The organization of this article is as follows:
In Sec.~\ref{sec:Dec}, we introduce 
the (de)construction of a massive gauge theory
with one (de)constructed extra dimension.
The mass spectrum in a vacuum of this model 
(``KK'' spectrum) is studied,
followed by the construction of 
the low energy effective theory
below the ``KK'' energy scale.
Here, we derive one of our main results 
that the field range of the 
``zero-mode'' of the gauge field in
the (de)constructed direction
is enhanced by the number of the lattice points.
Then we generalize the model to the case
with more (de)constructed extra dimensions.
To better understand 
the field range enhancement mechanism
in our model, 
we make a comparison 
with 
the corresponding model with 
continuous extra dimensions,
and make clear
the similarity and the difference.
In Sec.~\ref{sec:Model}
we construct a large field inflation model
based on our (de)constructed massive gauge theory.
The enhancement of the field range 
found in Sec.~\ref{sec:Dec}
is crucial for
achieving a trans-Planckian inflaton field excursion.
The rapid particle production mechanism
``\textit{excursion through KK modes}''
is extended to the case of ``KK'' modes of 
the (de)constructed extra dimensions,
and its cosmological
consequences are analyzed.
We conclude in Sec.~\ref{sec:Discussions} with 
summary and discussions.
In Appendix.~\ref{App:DFT}
we collect the relevant formulas
of discrete Fourier transformation
and also fix our convention.

\section{(De)construction of 
a massive gauge theory}\label{sec:Dec}
\subsection{The action}\label{subsec:S5}
We start with the
following 4D action 
with one (de)constructed extra dimension:
\begin{align}
S_{5}
=
\int d^4 x
\sum_{j=0}^{N-1}
\Biggl[
&- \frac{1}{4} F_{\mu\nu (j)}  F^{\mu\nu}_{(j)} 
\Biggr.
\nn\\
&+ 
\frac{f^2}{2}
D_\mu U_{(j,j+1)} D^{\mu} U_{(j,j+1)}^{\dagger}
+
V_S(U_{(j,j+1)},\theta_{(j)},\theta_{(j+1)})
\nn\\
&+
\Biggl.
D_\mu {\chi}_{(j)}^{\dagger} D^\mu \chi_{(j)} 
-
m^2 \chi_{(j)}^\dagger \chi_{(j)} 
+
f^2
\left(
\gamma \chi_{(j)}^\dagger U_{(j,j+1)}\chi_{(j+1)}
+ c.c.
\right)
+ \ldots
\Biggr]\,,
\nn\\
& \qquad \qquad \qquad \qquad (j=0,1,\cdots,N-1 \,\, \mathrm{mod} \,\, N).
\label{eq:decS5}
\end{align}
The field $U_{(j,j+1)}(x)$ takes value in $U(1)$
and can be parametrized as
\begin{equation}
U_{(j,j+1)}(x) = \exp \left[ i \frac{A_{(j,j+1)} (x)}{f} \right]\,.
\label{eq:Link}
\end{equation}
$A_{(j,j+1)} (x)$ is
an angular variable with
the period $2\pi f$.
The field $\chi_{(j)}$ is
a charged matter field
which for simplicity
we chose to be a scalar field.
The action \eqref{eq:decS5} 
has the product $U(1)^N$ gauge symmetry.
The gauge transformation generated by
$g_{(j)}(x) = e^{i g \alpha_{(j)}(x)}$ are given as
\begin{align}
A_{\mu (j)} (x) &\rightarrow A_{\mu (j)}(x)- \pa_\mu \alpha_{(j)}(x)\,,\\
U_{(j,j+1)} (x) &\rightarrow g_{(j)}^{-1}(x) U_{(j,j+1)}(x) g_{(j+1)}(x)\,,\\
\theta_{(j)} (x) &\rightarrow \theta_{(j)} (x) + \alpha_{(j)} (x)\,, \\
\chi_{(j)} (x) &\rightarrow g_{(j)}^{-1}(x) \chi_{(j)} (x) \,.
\end{align}
The covariant derivatives 
in \eqref{eq:decS5}
are defined as
\begin{align}
D_\mu U_{(j,j+1)}
&=
\pa_\mu U_{(j,j+1)} 
- i g A_{\mu (j)} U_{(j,j+1)} + i g U_{(j,j+1)} A_{\mu (j+1)} \,,
\label{eq:DU}\\
D_\mu \chi_{(j)}
&=
\pa_\mu \chi_{(j)} - i g A_{\mu (j)} \chi_{(j)} \,.
\label{eq:Dmuchi}
\end{align}
The action \eqref{eq:decS5}
may arise as 
a low energy EFT of strongly coupled gauge theory 
below the chiral symmetry breaking scale
\cite{ArkaniHamed:2001ca},
or as a low energy EFT of complex scalar fields
which acquire vacuum expectation values 
\cite{Hill:2000mu}.
In either case,
the field $A_{(j,j+1)}(x)$ in \eqref{eq:Link}
is the Nambu-Goldstone boson 
arising from $U(1)$ global symmetry breaking.
In the meantime, 
the (de)constructed dimension
has the same mathematical structure as
the lattice in lattice gauge theory.
In the terminology of lattice gauge theory,
$U_{(j,j+1)}(x)$ is the ``link variable''
on the lattice 
and $A_{(j,j+1)}(x)$ is identified with
the lattice gauge field
in the (de)constructed direction.
Following the terminology of 
lattice gauge theory,
we may define the ``lattice spacing'' $a$ 
from \eqref{eq:Link} as
\begin{equation}
a 
:= \frac{1}{g f}\,,
\label{eq:a}
\end{equation}
since in lattice gauge theory
the link variables are given as
\begin{equation}
U_{(j,j+1)} 
= 
\exp 
\left[
i g a A_{(j,j+1)} 
\right]
\,.
\label{eq:LinkLattice}
\end{equation}
We may also define the
``radius'' of the (de)constructed extra dimension $L$ by
\begin{equation}
L 
:= \frac{N a}{2\pi}  
= 
\frac{N}{2 \pi g f}
\,.
\label{eq:L}
\end{equation}
The field $\theta_{(j)}(x)$ 
is a (de)constructed
version of the Stueckelberg field
which allows us to have a
gauge invariant potential
for the gauge field in the
(de)constructed direction:
%
%
\begin{align}
V_S(A_{(j,j+1)},\theta_{(j)},\theta_{(j+1)})
&=
\sum_{p=0}^\infty
\tilde{V}_p
(e^{i \theta_{(j)}} U_{(j,j+1)} e^{-i \theta_{(j+1)}})^p 
+ c.c.
\nn\\
&=
\sum_{p=0}^\infty
\tilde{V}_p
\exp i p
\left[
\frac{A_{(j,j+1)}}{f}
- (\theta_{(j+1)} - \theta_{(j)})
\right]
+ c.c. \,.
\label{eq:PotFS}
\end{align}
Since $A_{(j,j+1)}$ is an angular variable
with the period $2\pi f$,
\eqref{eq:PotFS}
provides the Fourier series expansion
of the gauge invariant potential of $A_{(j,j+1)}$.
As in lattice field theory,
we have imposed
symmetry under the 
discrete translation
on the action \eqref{eq:decS5}:
\begin{align}
A_{\mu (j)} &\rightarrow A_{\mu (j+1)}\,,\\
U_{(j,j+1)} &\rightarrow U_{(j+1,j+2)}\,,\\
\theta_{(j)} &\rightarrow \theta_{(j+1)}\,,\\
\chi_{(j)} &\rightarrow \chi_{(j+1)} \,.
\end{align}
On the other hand,
there is no symmetry 
which mixes the continuous four space-time dimensions 
and the (de)constructed extra dimension, i.e.
there is no (1+4)D Lorentz symmetry. 
As a consequence, the 4D gauge fields do not need to
have Stueckelberg mass.
The lack of (1+4)D Lorentz symmetry 
gives rise to
interaction terms 
which do not exist in (1+4)D Lorentz symmetric models.
Some of these terms can be suppressed by
imposing the parity symmetry in the
(de)constructed extra dimension:
\begin{align}
j &\rightarrow N-j\quad (\mathrm{mod}\,\, N)\,,
\label{eq:Ptr1}\\
U_{(j,j+1)} &\rightarrow U_{(N-j,N-(j+1))} \,,
\label{eq:Ptr2}
\end{align}
where we have defined $U_{(j+1,j)}:= U^{\dagger}_{(j,j+1)}$.
The first line \eqref{eq:Ptr1} should be regarded as the change of
the label of the lattice points
so that in
\eqref{eq:Ptr2} the lattice point label
$j+1$ is transformed to $N-(j+1)$.
The parity symmetry forbids
terms which contain odd number of
the lattice counterparts of the derivative.
The parity symmetry also forbids the imaginary part of the 
parameter $\gamma$ in 
\eqref{eq:decS5} and
$\tilde{V}_p$ in \eqref{eq:PotFS}.
``$\ldots$'' in the action \eqref{eq:decS5}
represents 
higher dimensional operators
which will not be relevant at low energy
(more specifically, 
we will eventually be interested in the EFT 
below the ``KK'' energy scale $1/L$).

\subsection{The mass spectrum}\label{subsec:masses}

\subsubsection*{Gauge field 
(space-time components)}\label{subsubsec:massg}
The
mass-square matrix of the
gauge fields
in the vacuum
$U_{(j,j+1)} =1$ can be read off
from the action \eqref{eq:decS5}:
\begin{equation}
M_{g}^2 := g^2 f^2 K\,,
\label{eq:Mg2}
\end{equation}
where $K$ is the $N \times N$ matrix:
\begin{equation}
K:=
\left(
\begin{array}{cccccc}
2      & -1 & 0 & 0 & \cdots & -1\\
-1     & 2 & -1 & 0 & \cdots & 0\\
0 & -1 & 2 & -1 & \cdots & 0\\
\vdots &   &   & \ddots   & & \vdots\\
0   & \cdots &  &   & 2 & -1\\
-1 & \cdots & & & -1 & 2
\end{array}
\right)\,.
\label{eq:matK}
\end{equation}
The problem of finding the eigenvalues 
and the corresponding eigenstates
of the matrix $K$ given in \eqref{eq:matK}
is a familiar one 
which appears
in the system of coupled harmonic oscillators.
The fact that the
index $j$ is a periodic discrete variable
motivates us to use the
discrete Fourier transform
(our convention for the discrete Fourier transform
as well as relevant formulas
are summarized in Appendix~\ref{App:DFT}.):
\begin{equation}
A_{\mu (j)} 
= 
\frac{1}{\sqrt{N}}
\sum_{n= -\frac{N-1}{2}}^{\frac{N-1}{2}}
\tilde{A}_{\mu (n)} e^{i\frac{2\pi n j}{N}}\,,
\label{eq:Anodd}
\end{equation}
for odd $N$, and
\begin{equation}
A_{\mu (j)} 
= 
\frac{1}{\sqrt{N}}
\sum_{n= -\left(\frac{N}{2}-1\right)}^{\frac{N}{2}-1}
\tilde{A}_{\mu (n)} e^{i\frac{2\pi n j}{N}}
+
\frac{1}{\sqrt{N}}
\tilde{A}_{\mu (\frac{N}{2})}
(-)^j
\,,
\label{eq:Aneven}
\end{equation}
for even $N$.
Since $A_{\mu (j)}$ is a real variable,
the Fourier coefficients satisfy the relation
$\tilde{A}_{\mu (-n)}^\ast = \tilde{A}_{\mu (n)}$.
Note that $\tilde{A}_{\mu (0)}$,
and $\tilde{A}_{\mu (\frac{N}{2})}$ in 
the case of even $N$, are real.
The matrix multiplication of $K$ in \eqref{eq:matK}
to $A_{\mu (j)}$ in 
\eqref{eq:Anodd} or \eqref{eq:Aneven} gives
\begin{align}
\sum_{k=0}^{N-1}
K_{jk} {A}_{\mu (k)}
&=
2 {A}_{\mu (j)} -{A}_{\mu (j+1)}-{A}_{\mu (j-1)}
\nn\\
&=
\frac{1}{\sqrt{N}}
\sum_{n}
\tilde{A}_{\mu (n)} e^{i\frac{2\pi n j}{N}}
\left(
2 - e^{i \frac{2\pi n}{N}}-e^{-i \frac{2\pi n}{N}}
\right)
\nn\\
&=
\frac{1}{\sqrt{N}}
\sum_{n}
\tilde{A}_{\mu (n)} 
e^{i\frac{2\pi n j}{N}} 
4 \sin^2 \frac{\pi n}{N} \,,
\label{eq:KAn}
\end{align}
where the sum over $n$ should be taken
as in \eqref{eq:fnodd} or \eqref{eq:fneven}
depending on whether $N$ is odd or even.
From \eqref{eq:KAn} we find $N$
eigenvectors ${A}_{\mu (n)}^{m.e.v.}$
of the mass-square matrix
\eqref{eq:Mg2},
whose $j$-th component is given by
\begin{equation}
({A}_{\mu (n)}^{m.e.v.})_j 
= 
\tilde{A}_{\mu (n)} e^{i \frac{2 \pi n j}{N}}\,.
\label{eq:evg}
\end{equation}
The corresponding mass-square eigenvalue of
${A}_{\mu (n)}^{m.e.v.}$ is given by
\begin{equation}
M^2_{g(n)} =
4 g^2 f^2 \sin^2
\frac {\pi n}{N}\,.
\label{eq:Mg2eg}
\end{equation}
\eqref{eq:Mg2eg} 
provides
the lattice counterpart
of the KK mass spectrum
in the continuous extra dimension:
In the formal continuous limit
$N \rightarrow \infty$ and 
$a = 1/ gf \rightarrow 0$ with $L$ fixed,
the mass eigenvalues \eqref{eq:Mg2eg} reduce to
\begin{equation}
M^2_{g(n)}
\rightarrow
4 g^2 f^2 
\left(
\frac{\pi n}{N}
\right)^2
=
\left(
\frac{n}{L}
\right)^2
\,,
\label{eq:M2clim}
\end{equation}
when $n \ll N$.
\eqref{eq:M2clim} coincides with 
the KK mass spectrum from a compactification
on a continuous circle with radius $L$.
Note that the limit is formal, 
because
the parameters are constrained as
$g \lesssim 1$, $f \ll M_P$ 
in order for the model to be valid.
%
%
%

\subsubsection*{Gauge field 
((de)constructed extra 
dimensional component)}\label{subsubsec:massA}
To find the mass-square eigenvalues for
the field $A_{(j,j+1)}$,
as in the case of the gauge field (space-time components),
we consider the discrete Fourier transform
of $A_{(j,j+1)}$:
\begin{equation}
A_{(j,j+1)}
=
\frac{1}{\sqrt{N}}
\sum_{n}
\tilde{A}_{(n)}
e^{i\frac{2\pi n j}{N}}\, ,
\label{eq:ADFT}
\end{equation}
where the sum over $n$ is taken
as in \eqref{eq:fnodd} for odd $N$ 
and as in \eqref{eq:fneven} for even $N$.

We impose an analogue of the Lorenz gauge condition
including the (de)constructed extra-dimension:
\begin{equation}
\pa_\mu \tilde{A}_{(n)}^\mu 
-
i M_{g(n)} \tilde{A}_{(n)}
= 0 \,,
\label{eq:LorenzG}
\end{equation}
where $M_{g(n)}$ is the positive square root of 
$M_{g(n)}^2$ in \eqref{eq:Mg2eg}.
%
In this gauge, 
apart from the Stueckelberg mass term
arising from the potential
\eqref{eq:PotFS},
the field $\tilde{A}_{(n)}$
acquires 
the ``KK'' mass term.
This ``KK'' contribution 
$M_{\tilde{A}(n)}^{2\, KK}$
to the mass-square of
the field $\tilde{A}_{n}$
is given as
\begin{equation}
M_{\tilde{A}(n)}^{2\, KK} =
4 g^2 f^2 \sin^2
\frac {\pi n}{N}\,.
\label{eq:MA2eg}
\end{equation}

\subsubsection*{Matter field}\label{subsubsec:masschi}
Below the ``KK'' energy scale
$1/L$,
an appropriate EFT description
is obtained by
integrating out fields which have 
mass greater than $1/L$.
Therefore we 
replace all $A_{(j,j+1)}$ to its 
``zero-mode'',
i.e.
the $n=0$ term in \eqref{eq:ADFT}:
\begin{equation}
A_{(j,j+1)} 
\Bigl|_{\tilde{A}_{(n)} = 0\,\, \mathrm{except}\,\, n=0}
= \frac{1}{\sqrt{N}} \tilde{A}_{(0)} 
:= \frac{1}{\sqrt{N}} A\,,
\label{eq:repAzero}
\end{equation}
where the Fourier coefficient
$A = \tilde{A}_{(0)}$ is expressed as
(see Appendix~\ref{App:DFT})
\begin{equation}
A =
\tilde{A}_{(0)}
=
\frac{1}{\sqrt{N}}
\sum_{j=0}^{N-1}
A_{(j,j+1)}\,.
\label{eq:A0}
\end{equation}
The part of the action \eqref{eq:decS5}
which involves the matter fields 
$\chi_{(j)}(x)$ 
can be rewritten as
\begin{align}
S_{\chi}
=
\int d^4x
\Bigl[
&D_\mu {\chi}_{(j)}^{\dagger} 
D^\mu \chi_{(j)} 
-
(m^2 - 2 \gamma f^2)
\chi_{(j)}^\dagger \chi_{(j)}
\nn\\
&-
\gamma f^2
(U_{(j,j+1)}\chi_{(j+1)} -\chi_{(j)})^\dagger
(U_{(j,j+1)}\chi_{(j+1)} -\chi_{(j)})
\Bigr]\,.
\label{eq:Schi}
\end{align}
From \eqref{eq:Schi}
we observe that the mass of the field $\chi_{(j)}$
depends on 
the expectation value of $A$.
The mass-square matrix for the field $\chi_{(j)}$
is given as
\begin{equation}
M_{\chi}^2 (A) = 
(m^2 - 2\gamma f^2)
+
\gamma f^2 K(A) \,,
\label{eq:Mchi2}
\end{equation}
where
\begin{equation}
K(A):=
\left(
\begin{array}{cccccc}
2      & -e^{i\frac{A}{\sqrt{N}f}} & 0 & 0 & \cdots & -e^{-i\frac{A}{\sqrt{N}f}}\\
-e^{-i\frac{A}{\sqrt{N}f}}      & 2 & -e^{i\frac{A}{\sqrt{N}f}}  & 0 & \cdots & 0\\
0 & -e^{-i\frac{A}{\sqrt{N}f}}  & 2 & -e^{i\frac{A}{\sqrt{N}f}}  & \cdots & 0\\
\vdots &   &   & \ddots   & & \vdots\\
0   & \cdots &  &   & 2 & -e^{i\frac{A}{\sqrt{N}f}} \\
-e^{i\frac{A}{\sqrt{N}f}} & \cdots & & & -e^{-i\frac{A}{\sqrt{N}f}}  & 2
\end{array}
\right)\,.
\label{eq:KA}
\end{equation}
As before, 
to find the mass-square eigenvalues, 
we consider
the discrete Fourier transform of $\chi_{(j)}$:
\begin{equation}
\chi_{(j)} 
= 
\frac{1}{\sqrt{N}}
\sum_{n}
\tilde{\chi}_{(n)} e^{i\frac{2\pi n j}{N}}\,,
\label{eq:chiell}
\end{equation}
where the sum over $n$ is taken
as in \eqref{eq:fnodd} or \eqref{eq:fneven},
according to whether $N$ is odd or even.
Multiplying the mass-square matrix to
$\chi_{(j)}$,
we obtain
\begin{align}
&\sum_{k=0}^{N-1}
K(A)_{jk} \chi_{(k)}
\nn\\
&=
2 {\chi}_{(j)} 
- e^{i\frac{A}{\sqrt{N}f}}  {\chi}_{(j+1)}
- e^{-i\frac{A}{\sqrt{N}f}}  {\chi}_{(j-1)}
\nn\\
&=
\frac{1}{\sqrt{N}}
\sum_{n}
\tilde{\chi}_{(n)} 
e^{i \frac{2\pi n j}{N}}
\left(
2 
- 
\exp 
\left[
i 
\left(
\frac{2\pi n}{N} 
+ \frac{A}{\sqrt{N}f}
\right)
\right]
- 
\exp 
\left[
-i 
\left(
\frac{2\pi n}{N}  
+ \frac{A}{\sqrt{N}f}
\right)
\right]
\right)
\nn\\
&=
\frac{1}{\sqrt{N}}
\sum_{n}
\tilde{\chi}_{(n)} 
e^{i\frac{2\pi n j}{N}} 
4 \sin^2 
\left[
\frac{1}{2 \sqrt{N} f}
\left( 
A + \frac{2\pi f n}{\sqrt{N}}
\right)
\right]
\nn\\
:&=
\frac{1}{\sqrt{N}}
\sum_{n}
\tilde{\chi}_{(n)} 
e^{i\frac{2\pi n j}{N}} 
4 \sin^2 
\left[
\frac{1}{2 F}
\left( 
A + {2\pi f_4 n}
\right)
\right]
\,,
\label{eq:KAchi}
\end{align}
where
\begin{equation}
F:= \sqrt{N} f \,,
\label{eq:F}
\end{equation}
and
\begin{equation}
f_4 := \frac{f}{\sqrt{N}} \,.
\label{eq:f4}
\end{equation}
From \eqref{eq:KAchi} we find the
eigenvectors ${\chi}_{(n)}^{m.e.v.}$
of the mass-square matrix \eqref{eq:Mchi2},
whose $j$-th component is given by
\begin{equation}
(\tilde{\chi}_{(n)}^{m.e.v})_j 
= 
\tilde{\chi}_{(n)} e^{i \frac{2 \pi n j}{N}}\,.
\label{eq:evchi}
\end{equation}
The eigenvector ${\chi}_{(n)}^{m.e.v.}$ has the
eigenvalue $M^2_{\chi(n)}(A)$ given by
\begin{equation}
M^2_{\chi(n)}(A)
= 
(m^2 - 2\gamma f^2) 
+
M^{2\, KK}_{\chi(n)}(A) 
\,,
\label{eq:Mchi2n}
\end{equation}
where the ``KK'' mass spectrum of $\chi$ is given by
\begin{equation}
M^{2\,KK}_{\chi(n)}(A)
=
4 \gamma f^2 \sin^2 
\left( \frac{A 
+ 2 \pi f_4 n
}{2 F}
\right)\, .
\label{eq:Mchi2nKK}
\end{equation}
Unlike the real extra dimension
whose radius appears the same 
for all the fields propagating in it,
in (de)construction 
the ``KK'' mass spectrum of $\chi$
given in \eqref{eq:Mchi2nKK}
need not coincide with that of the gauge field 
even when $A=0$.
The fields canonically normalized in
$(1+3)$D
can have different coefficients for 
their ``kinetic'' term,
i.e. the lattice counterpart of the kinetic term,
in the (de)constructed extra dimension.
In other words,
different fields propagate 
in the (de)constructed extra dimension differently.
This lack of universality 
in the (de)constructed extra dimension
leads to the 
absence of Lorentz symmetry between
the $(1+3)$D space-time dimensions
and the extra dimensions
even in the formal continuum limit $N \rightarrow \infty$.
This breaking of universality/Lorentz symmetry
in/involving the (de)constructed extra dimension
is parametrized by the real parameter $\gamma$.
The universality in the (de)constructed extra dimension
recovers at $\gamma = g^2$,
and only in this case their ``KK'' mass spectra coincide.


\subsection{The low energy effective action
below KK energy scale}\label{subsec:S4}
Now we would like to have 
the low energy effective action
of \eqref{eq:decS5} which is
suitable for describing the physics
below the ``KK'' energy scale $1/L$.
After integrating out the fields
whose mass 
is above the ``KK'' energy scale $1/L$,
we obtain the
effective action $S_4$:
\begin{align}
S_4
=
\int d^4 x
\Biggl[
- 
&
\frac{1}{4} F_{\mu\nu}  F^{\mu\nu}
\Biggr.
+ 
\frac{1}{2}
\pa_\mu A \pa^{\mu} A
+
V_S^{(4)}(A)
\nn\\
&+
\Biggl.
\sum_{n}
\left(
D^{(4)}_\mu \tilde{\chi}_{(n)}^{\dagger} D^{(4)\mu} \tilde{\chi}_{(n)}
-
\tilde{\chi}_{(n)}^\dagger 
M^2_{\chi(n)}(A)
\tilde{\chi}_{(n)}
\right)
\Biggr]\,.
\label{eq:decS4}
\end{align}
Here, the 4D gauge field $A_\mu$
is the ``zero-mode''
of the (de)constructed extra dimension:
\begin{equation}
A_{\mu}
:=
\tilde{A}_{\mu(0)}
=
\frac{1}{\sqrt{N}}
\sum_{j=0}^{N-1}
A_{\mu (j)}\,.
\label{eq:Amu0}
\end{equation}
The covariant derivative 
for the unbroken diagonal $U(1)$
is given by
\begin{equation}
D_\mu^{(4)} \tilde{\chi}_{(n)}
=
\pa_\mu \tilde{\chi}_{(n)}
- i g_4 A_\mu \tilde{\chi}_{(n)}\,,
\label{eq:Dmutchi}
\end{equation}
where the gauge coupling $g_4$
for the unbroken diagonal $U(1)$
is given by\footnote{%
Those who are familiar with the
Weak Gravity Conjecture (WGC) \cite{ArkaniHamed:2006dz}
might worry that by taking $N$ large,
$g_4$ becomes too small so that 
the model is in tension with the conjecture
that gravity is the weakest force.
However, it has been shown that when
WGC is respected at high energy,
it is not violated by Higgs mechanism
\cite{Saraswat:2016eaz,Furuuchi:2017upe,Heidenreich:2017sim}.
In the current case, if $g$ is not too small,
the model would not have a tension with WGC.}
\begin{equation}
g_4 := \frac{g}{\sqrt{N}} \,.
\label{eq:g4}
\end{equation}
Since the mass of the charged fields
$\tilde{\chi}_{(n)}$
(the range of $n$ is as in
\eqref{eq:fnodd} or \eqref{eq:fneven}
depending on whether $N$ is odd or even)
depend on
the value of the field $A$,
we do not know which of the field $\tilde{\chi}_{(n)}$
becomes light before knowing the value of $A$.
Hence we keep all $\tilde{\chi}_{(n)}$
in the low energy effective action \eqref{eq:decS4}.

The (de)constructed version 
of the Stueckelberg field $\theta_{(j)}$
is expanded in discrete Fourier modes as
\begin{equation}
\theta_{(j)}
=
\frac{1}{\sqrt{N}}
\sum_{n}
\tilde{\theta}_{(n)}
e^{i\frac{2\pi n j}{N}}
+ \frac{2\pi w j}{N}\, .
\label{eq:thetamode}
\end{equation}
Since $\theta_{(j)}$ ($j=0,1,\cdots N-1$ \,\, $\mathrm{mod}\,\, N$)
are angular variables: $\theta_{(j)} \sim \theta_{(j)} + 2\pi$,
they can have a ``winding number'' 
$w$ which is an integer.\footnote{%
$A_{(j,j+1)}$ could also have
a winding number, however
in the current gauge \eqref{eq:LorenzG}
the winding mode has large energy and
decouples at low energy.}
The Stueckelberg potential 
\eqref{eq:PotFS}
leads to the potential
$V_S^{(4)}(A)$ for the ``zero-mode'' $A$:
\begin{align}
V_{S}^{(4)}(A)
&=
2N 
\sum_{p=0}^\infty
\tilde{V}_p 
\cos 
\left[
p 
\left(
\frac{A}{\sqrt{N} f}
-
\frac{2\pi w}{N}
\right)
\right]
\nn\\
&=
2N 
\sum_{p=0}^\infty
\tilde{V}_p 
\cos 
\left[
\frac{p(
{A}
-
{2\pi f_4 w}
)}{F}
\right]
\,.
\label{eq:VS4}
\end{align}
%
\eqref{eq:VS4}
is one of our main results.
Compared with the period $2\pi f$ 
of the original Stueckelberg potential
\eqref{eq:PotFS},
the period of the potential \eqref{eq:VS4}
enhanced by a factor $\sqrt{N}$.
By taking $N$ large,
the model may allow a large field 
excursion.
We will discuss the application of this mechanism 
in cosmology in Sect.~\ref{sec:Model}.
In addition, the height of the potential $V_S^{(4)}$
is enhanced by a factor $N$
compared with the functional form of the original
Stueckelberg potential \eqref{eq:PotFS}.

In addition to the action
\eqref{eq:decS4},
we should include the following
``Wilson loop operator''
\begin{align}
W(A) :&=
\prod_{j=0}^{N-1}
U_{(j, j+1)}
=
\exp 
\left[
i 
\frac{%
\sum_{j=0}^{N-1}
A_{(j,j+1)}%
}{f}
\right]
\nn\\
&=
\exp 
\left[
i 
\frac{%
\sqrt{N} \tilde{A}_{(0)}%
}{f}
\right]
=
\exp 
\left[
i 
\frac{%
{A}%
}{f_4}
\right]
\,,
\label{eq:WL}
\end{align}
where $f_4$ is defined in \eqref{eq:f4}.
The Wilson loop operator
is generated 
in the 1-PI effective action
at 1-loop level as
\begin{equation}
c_{WL}
f^2 \Lambda^2
\left(
\frac{\gamma f^2}{\Lambda^2}
\right)^N
W(A)
+ c.c. \,,
\label{eq:WLinS4}
\end{equation}
where $\Lambda = 4\pi f$
is the UV energy scale 
at which the perturbative expansion of the EFT
\eqref{eq:decS4} breaks down.
The natural magnitude of the coefficient $c_{WL}$ is 
$\Ord(1)$.
$c_{WL}$ is real as we have imposed a symmetry
under the parity transformation
\eqref{eq:Ptr1}, \eqref{eq:Ptr2}.
The Wilson loop operator
\eqref{eq:WL} contributes to the
potential of the ``zero-mode'' $A$:
\begin{equation}
V_{WL}(A)
=
2 c_{WL}
f^2 \Lambda^2
\left(
\frac{\gamma f^2}{\Lambda^2}
\right)^N
\cos
\left(
\frac{A}{f_4}
\right)\,.
\label{eq:WLpot}
\end{equation}

It is important to notice that the ``KK'' mass term
of the charged fields $\tilde{\chi}_{(n)}$
summed over the ``KK'' modes $n$:
\begin{equation}
\sum_{n}
\tilde{\chi}_{(n)}^\dagger 
M^{2\,KK}_{\chi(n)}(A)
\tilde{\chi}_{(n)}\,,
\label{eq:KKmassterm}
\end{equation}
is invariant under the following transformation:
\begin{align}
A 
&\rightarrow 
A + 2\pi f_4 \,,
\label{eq:gt1}\\
\tilde{\chi}_{(n)} 
& \rightarrow 
\tilde{\chi}_{(n+1)}\,.
\label{eq:gt2}
\end{align}
The invariance 
of the mass term \eqref{eq:KKmassterm}
under the transformation 
\eqref{eq:gt1}, \eqref{eq:gt2}
originates from the $U(1)^N$ gauge symmetry.
To understand it, first notice that
the gauge parameters $\alpha_{(j)}$
($j=0,1,\cdots,N-1$ $\mathrm{mod}\,\,N$)
are angular variables:
\begin{equation}
\alpha_{(j)} \sim \alpha_{(j)} + \frac{2\pi}{g}\,.
\label{eq:aang}
\end{equation}
The following choice of the gauge parameter
is a legitimate gauge transformation
due to \eqref{eq:aang}:
\begin{equation}
\alpha_{(j)}
=
\frac{2\pi j}{g N} \,.
\label{eq:Apgp}
\end{equation}
The gauge transformation 
\eqref{eq:Apgp}
leads to the transformation
\eqref{eq:gt1}, \eqref{eq:gt2}.
Note that 
$f_4$ appearing in \eqref{eq:gt1}
is the same as the one appearing in the 
Wilson loop potential \eqref{eq:WLpot},
because its periodicity 
also follows from the same gauge transformation
by \eqref{eq:Apgp}
and its gauge invariance.
Also note that the period $f_4$ becomes smaller 
as we take $N$ large
(provided that we do not scale $f$ with $N$).

From \eqref{eq:Mchi2n},
we observe that
in the case 
$m^2 - 2 \gamma f^2 =0$,
the field $\tilde{\chi}_{(n)}$
becomes massless
when $A + 2\pi f_4 n =0$,\footnote{%
More precisely, when $A + 2\pi f_4 n$  
equals integer multiple of $2\pi F$.
However, we will be mostly interested
in the field range of $A$ 
much smaller than $2\pi F$.}
where $n$ is an integer
in the range specified in
\eqref{eq:fnodd} or \eqref{eq:fneven}.
This has interesting consequences in cosmology,
which we explore in Sec.~\ref{sec:Model}.
Notice that small $m^2 - 2 \gamma f^2$ is natural in
the sense of 't Hooft \cite{tHooft:1979rat},
as it can be protected by the
approximate shift symmetry
in the action \eqref{eq:decS5}
(which may be clearer from \eqref{eq:Schi}):
\begin{equation}
\chi_{(j)} \rightarrow \chi_{(j)} + c 
\qquad \mbox{for all}\,\, j \,,
\label{eq:shiftsymm}
\end{equation}
where $c$ is a complex constant.
The shift symmetry is a good symmetry
when the gauge coupling $g$ and 
the coupling $\gamma$ are also small.
Note that these parameters are 
related as $\gamma = g^2$
at the point of physical interest 
where 
the universality of the (de)constructed
extra dimensions recovers, 
as explained in the end of 
Sec.~\ref{subsec:masses}.



\subsection{Generalization to higher 
(de)constructed extra dimensions}\label{subsec:HD}
Below, 
we generalize 
the model to have more (de)constructed extra dimensions.
Let us consider
a $d$-dimensional periodic lattice
(a lattice on a $d$-dimensional torus)
with $N_{I}$ 
(${I} = 1,2,\cdots,d$)
lattice points
in the ${I}$-th direction.
The action 
we consider is a straightforward
generalization of 
\eqref{eq:decS5}:
\begin{align}
S_{(4+d)}
=
\int d^4 x
\sum_{\vec{j}}
\Biggl[
&- \frac{1}{4} F_{\mu\nu (\vec{j})}  F^{\mu\nu}_{(\vec{j})} 
+
D_\mu {\chi}_{(\vec{j})}^{\dagger} D^\mu \chi_{(\vec{j})} 
-
m^2 \chi_{(\vec{j})}^\dagger \chi_{(\vec{j})} 
\Biggr.
\nn\\
&+ 
\sum_{I=1}^d
\Biggl\{\Biggr.
\frac{f_I^2}{2}
D_\mu U_{(\vec{j},\vec{j}+\vec{e}_I)}^I D^{\mu} U_{(\vec{j},\vec{j}+\vec{e}_I)}^{I\dagger}
+
V_S^{(4+d)}(A^I_{(\vec{j},\vec{j}+\vec{e}_I)},\theta_{(\vec{j})},\theta_{(\vec{j}+\vec{e}_I)})
\nn\\
&
\Biggl.
\qquad \qquad
+
f_I^2
\left(
\gamma_I \chi_{(\vec{j})}^\dagger U^I_{(\vec{j},\vec{j}+\vec{e}_I)}\chi_{(\vec{j}+\vec{e}_I)}
+ c.c.
\right)
\Biggl.\Biggr\}
+ \ldots
\Biggr]\,,
\nn\\
& \qquad \qquad \qquad \qquad (j_I=0,1,\cdots,N_I-1 \,\, \mathrm{mod} \,\, N_I).
\label{eq:decS4pd}
\end{align}
The $I$-th component of $\vec{n}$
and $\vec{j}$
are denoted as 
$n_I$ and $j_I$, respectively.
The link variable can be parametrized as
\begin{equation}
U_{(\vec{j},\vec{j}+\vec{e}_{I})}^{{I}}
=
\exp
\left[
i \frac{A_{(\vec{j},\vec{j}+\vec{e}_{I})}^{I}}{f_{I}}
\right]\,,
\label{eq:Ualpha}
\end{equation}
where the $d$-dimensional vector $\vec{j}$ 
parametrizes the lattice points
and $\vec{e}_{I}$ is a vector 
whose ${J}$-th component is given by $\delta_{{I} {J}}$.
There are $d$ components of
the gauge field in the (de)constructed directions, 
$A^I_{(\vec{j},\vec{j}+\vec{e}_{I})}$
($I = 1,2,\cdots,d$).
The Stueckelberg potential is given as
\begin{align}
&V_S^{(4+d)}
(
A_{(\vec{j},\vec{j}+\vec{e}_{I})}^I,
\theta_{(\vec{j})},
\theta_{(\vec{j}+\vec{e}_{I})}
)
\nn\\
=&
\sum_{p_1=0}^\infty
\sum_{p_2=0}^\infty
\cdots
\sum_{p_d=0}^\infty
\tilde{V}_S(p_1,p_2,\cdots,p_d)
\exp  
\left[
i
\sum_{I=1}^{d}
p_I
\left(
\frac{A_{(\vec{j},\vec{j}+\vec{e}_I)}^I}{f_I}
- (\theta_{(\vec{j}+\vec{e}_I)} - \theta_{(\vec{j})})
\right)
\right]
+ c.c. \,.
\label{eq:PotFSHD}
\end{align}
As before, we consider the discrete Fourier transform
\begin{equation}
A_{(\vec{j},\vec{j}+\vec{e}_I)}^I
=
\frac{1}{\prod_{{I}=1}^d N_{I}^{{1}/{2}}}
\sum_{\vec{n}}
\tilde{A}_{(\vec{n})}^I
e^{i \sum_{I=1}^{d} \frac{2\pi n_I j_I}{N_I}}\,.
\label{eq:ADFTHD}
\end{equation}
Here, the sum over $\vec{n}$,
we apply our discrete Fourier transform
convention
\eqref{eq:fnodd} or \eqref{eq:fneven} 
to each component $n_I$.

We will eventually be 
interested in one of the components of $A^I$
in the application to 
a single field inflation model
to be discussed in Sect.~\ref{sec:Model}.
To have such a single field inflation model,
we choose a potential \eqref{eq:PotFSHD}
in such a way that except for ${I} =1$
all the components have mass larger 
than the 
lowest of the ``KK'' energy scales
$1/L_I := 2\pi g f_I / N_I$.
Then below the lowest ``KK'' energy scale
we can set 
$A_{(\vec{j},\vec{j}+\vec{e}_I)}^I =0$
except for ${I} =1$
in the potential \eqref{eq:PotFSHD}.
We can also set $\theta_{(\vec{j})}=0$,
as the winding number can be set to zero 
by a choice of gauge.
We regard this 
$A_{(\vec{j},\vec{j}+\vec{e}_I)}^I =0$ ($I\ne 1$),
$\theta_{(\vec{j})}=0$
slice of 
$V_S^{(4+d)}(A^I_{(\vec{j},\vec{j}+\vec{e}_I)},\theta_{(\vec{j})},\theta_{(\vec{j}+\vec{e}_{I})})$
as a function of $A_{(\vec{j},\vec{j}+\vec{e}_{1})}^1$
which we call $V_1(A_{(\vec{j},\vec{j}+\vec{e}_{1})}^1)$,
and consider its Fourier series expansion:
\begin{align}
V_1(A_{(\vec{j},\vec{j}+\vec{e}_1)}^{1})
:&=
V_S^{(4+d)}
\left( A_{(\vec{j},\vec{j}+\vec{e}_1)}^{1},
A_{(\vec{j},\vec{j}+\vec{e}_{I})}^I = 0\,\, (I\ne 1),
\theta_{(\vec{j})}=0
\right)
\nn\\
\nn\\
&=
\sum_{p_1=0}^\infty
\tilde{V}_1(p_1)
\exp  
\left[
i
p_1
\left(
\frac{A_{(\vec{j},\vec{j}+\vec{e}_1)}^1}{f_1}
\right)
\right]
+ c.c. 
\,.
\label{eq:AIset0}
\end{align}
We can further set 
$\tilde{A}_{(\vec{n})}^1 = 0$
except for the ``zero-mode'' 
$\vec{n}=\vec{0}$,
which we call $A$:
\begin{equation}
A
:=
\tilde{A}_{(\vec{0})}^{1}
=
\frac{1}{\prod_{{I}=1}^d N_{I}^{1/2}}
\sum_{\vec{j}}
A_{(\vec{j},\vec{j}+\vec{e}_{1})}^{1}\,,
\label{eq:A0d}
\end{equation}
\begin{equation}
A_{(\vec{j},\vec{j}+\vec{e}_1)}^{1}
\Bigl|_{\tilde{A}^{1}_{(\vec{n})} =0 
\,\, \mathrm{except}\,\, \vec{n}=\vec{0}}
=
\frac{1}{\prod_{{I}=1}^d N_{I}^{1/2}}
\tilde{A}_{(\vec{0})}^{1} 
=
\frac{1}{\prod_{{I}=1}^d N_{I}^{1/2}}
A
\,.
\label{eq:AI0}
\end{equation}
Substituting \eqref{eq:AI0}
to
\eqref{eq:AIset0}, 
we obtain
\begin{align}
V_S^{(4)}(A) 
:&=
\sum_{\vec{j}}
V_1
\left(
A_{(\vec{j},\vec{j}+\vec{e}_1)}^{1}
\Bigl|_{\tilde{A}^{1}_{(\vec{n})} =0 
\,\, \mathrm{except}\,\, \vec{n}=\vec{0}}
\right)
\nn\\
&=
\left( \prod_{{I}=1}^d N_{I} \right)
\sum_{p_1}
\tilde{V}_1(p_1)
\exp  
\left[
i
p_1
\left(
\frac{A}{F}
\right)
\right]
+ c.c. 
\,,
\label{eq:A1nset0}
\end{align}
where we have defined
\begin{equation}
F := f_1 \prod_{{I}=1}^d N_{I}^{1/2}
\,.
\label{eq:FHD}
\end{equation}
We observe that
compared with
the period $2\pi f_1$
of the 
original potential \eqref{eq:AIset0},
the period $2\pi F$ of the potential \eqref{eq:A1nset0}
is enhanced
by a factor 
$\prod_{{I}=1}^d N_{I}^{1/2}$.
The height of 
the potential \eqref{eq:A1nset0}
is also enhanced 
compared with that 
in the original potential \eqref{eq:AIset0}
by a factor $\prod_{{I}=1}^d N_{I}$.

Next we introduce 
the
Wilson loop operator
wrapping around the $I=1$ direction:
\begin{align}
W_1(A)
:&=
\exp 
\left[
i
\sum_{j=0}^{N_1-1}
\frac{A_{(j \vec{e}_1, (j+1)\vec{e}_1)}^1}{f_1}
\right]
=
\exp 
\left[
i
\frac{N_1 A}{f_1 \prod_{{I}=1}^d N_{I}^{1/2}}
\right]
\nn\\
&=
\exp 
\left[
i
\frac{A}{f_4}
\right]
\,,
\label{eq:WLHD}
\end{align}
where
\begin{equation}
f_4:=
\frac{f_1 \prod_{I=1}^d N_I^{1/2}}{N_1} \,.
\label{eq:f4decHD}
\end{equation}
Thus the Wilson-loop potential
has a period $2\pi f_4$.

Next we study charged scalar fields
$\chi_{(\vec{j})}$.
The discrete Fourier transform of
the charged field $\chi_{(\vec{j})}$
is given by
\begin{equation}
\chi_{(\vec{j})}
=
\frac{1}{\prod_{{I}=1}^d N_{I}^{1/2}}
\sum_{\vec{n}}
\tilde{\chi}_{(\vec{n})}
e^{i \sum_{I=1}^{d} \frac{2\pi n_I j_I}{N_I}}\,.
\label{eq:chiDFTHD}
\end{equation}
Below the ``KK'' energy scale, 
we can set 
$\tilde{\chi}_{(\vec{n})} = 0$
except for
$\vec{n} = (n_1,0,\cdots,0)$.
Then 
$N_1$ eigenvalues
of the mass-square matrix of $\chi$ 
are given as
\begin{equation}
M^2_{\chi(n_1)}
=
(m^2 - 2 \sum_{I=1}^d \gamma_I f_I^2)
+
M^{2\,KK}_{\chi(n_1)} \,,
\label{Mchi2HD}
\end{equation}
where the ``KK'' contribution to the mass-square
is given by
\begin{equation}
M^{2\,KK}_{\chi(n_1)}
=
4
\gamma_1 f_1^2
\sin^2
\left(
\frac{A + 2 \pi f_4 n_1}{2 F}
\right)\,.
\label{eq:Mchi2HDKK}
\end{equation}

\subsection{Comparison with
continuous extra dimensions}\label{subsec:cont}
It will be instructive to make a comparison
with 
a $(4+d)$D massive gauge theory 
compactified on a $d$-dimensional torus. 
For the purpose of illustration,
we only write down the terms
in the action of the massive gauge theory
which are 
relevant for understanding the correspondence:
\begin{equation}
S_{(4+d)}^c = \int d^4x \int d^dy 
\left[
\frac{1}{2} \pa_\mu A^1(x,\vec{y}) \pa^\mu A^1(x,\vec{y})
-
2V_{(4+d)}
\cos
\left(
\frac{A^1}{f^c}
\right)
\right]
+
\ldots
\,,
\label{eq:S4d}
\end{equation}
where $x$ is the 4-coordinate of uncompactified space-time 
and
$\vec{y}$ is the coordinate vector on the 
$d$-dimensional torus.
$A^1(x,\vec{y})$
is the $1$st of the $d$ components 
of the massive gauge field in the compactified directions.
$f^c$
is a constant with mass-dimension 
$(d+2)/2$
and 
$V_{(4+d)}$ is a constant with mass-dimension 
$4+d$.
The potential of the massive gauge field
in \eqref{eq:S4d} 
is to be compared with
that in 
\eqref{eq:AIset0} 
with all 
$\tilde{V}_1(p_1)$ set to zero
except for $p_1=1$:
\begin{align}
V_1(A_{(\vec{j},\vec{j}+\vec{e}_1)}^{1})
&=
\tilde{V}_1(p_1=1)
\exp  
\left[
i
\left(
\frac{A_{(\vec{j},\vec{j}+\vec{e}_1)}^1}{f_1}
\right)
\right]
+ c.c. 
\nn\\
&=
2 \tilde{V}_1(p_1=1) 
\cos 
\left(
\frac{A_{(\vec{j},\vec{j}+\vec{e}_1)}^1}{f_1}
\right)
\,.
\label{eq:tV1p1}
\end{align}
Consequently 
\eqref{eq:A1nset0}
becomes
\begin{align}
V_S^{(4)}(A)
&=
\left(
\prod_{I=1}^{d} N_I
\right)
2 \tilde{V}_1(p_1=1) 
\cos 
\left(
\frac{A}{F}
\right)
\nn\\
&=
2V_4
\cos 
\left(
\frac{A}{F}
\right)
\,,
\label{eq:VS4p11}
\end{align}
where
\begin{equation}
V_4 := 
\left(
\prod_{I=1}^{d} N_I
\right)
\tilde{V}_1(p_1=1) \,.
\label{eq:decV4}
\end{equation}

For simplicity, we assume that
the compactification radii
to be the same $L$ in all compactified directions.
With a bit of abuse of notation,
we use the same symbols 
for both the (de)constructed model
and the model with continuous extra dimensions
when the correspondence of the quantity is clear,
like $L$ here, $A$, $g_4$, $f_4$ and $F$,
and $V_4$ to be introduced shortly.
The correspondence between 
the model with the (de)constructed extra dimensions
and that with the continuous extra dimensions 
are summarized in Table.~\ref{table:DeCont}.
Then the $(4+d)$D (reduced) Planck scale 
and the $4$D (reduced) Planck scale $M_P$
are related as
\begin{equation}
M_P^2 = (2\pi L)^d M_{(4+d)}^{2+d} \,.
\label{eq:M4M4d}
\end{equation}
It is natural to assume that
$f^c$ 
is sub-Planckian:\footnote{Here we are not claiming
that $(f^c)^{\frac{2}{d+2}} > M_{(4+d)}$ 
is fundamentally inconsistent.
It has been proposed that
if an EFT can be embedded 
in string theory,
the field range is restricted by the Planck scale:
$(f^c)^{\frac{2}{d+2}} < M_{(4+d)}$
(the distance conjecture \cite{Ooguri:2006in}).
Our viewpoint is that firstly this proposal is still a conjecture, 
and secondly even if the field domain in the original EFT
is restricted to be sub-Planckian,
there are mechanisms to obtain 
a trans-Planckian field range at low energy,
such as axion monodromy 
\cite{Kim:2004rp,Silverstein:2008sg,McAllister:2008hb}
or as we discuss in this article.
There is no obvious reason why the distance conjecture
should constrain such enhancements of effective field range.}
\begin{equation}
f^c \lesssim M_{(4+d)}^{\frac{d+2}{2}} \,.
\label{eq:FM4d}
\end{equation}

The Fourier expansion
of the field $A^1(x,\vec{y})$
is given as
\begin{equation}
A^1(x,\vec{y})
=
\frac{1}{(2\pi L)^{\frac{d}{2}}}
\sum_{\vec{n}}
\tilde{A^1}(x,\vec{n}) e^{i \frac{\vec{n} \cdot \vec{y}}{L}}\,.
\label{eq:FText}
\end{equation}
The 
Fourier coefficients
are given as
\begin{equation}
\tilde{A^1}(x,\vec{n})
=
\frac{1}{(2\pi L)^{\frac{d}{2}}}
\int d^{d}y\,
\tilde{A^1}(x,\vec{y}) e^{-i \frac{\vec{n} \cdot \vec{y}}{L}}\,.
\label{eq:FT4}
\end{equation}
From \eqref{eq:FT4}, when 
$\tilde{A^1}(x,\vec{n})$
is set to zero except for the zero-mode $\vec{n}=\vec{0}$,
\begin{equation}
A^1(x,\vec{y})
\Bigl|_{\tilde{A^1}(x,\vec{n})=0
\,\,
\mathrm{except}
\,\, \vec{n}=\vec{0}} = 
\frac{1}{(2\pi L)^{\frac{d}{2}}}
\tilde{A^1}(x,\vec{n}=\vec{0})\,,
\label{eq:zerod}
\end{equation}
we obtain 
$4$D action $S_4$
at the classical level:
\begin{equation}
S_{4} = \int d^4 x 
\left[
\frac{1}{2} \pa_\mu {A} \pa^\mu {A}
-
2V_4
\cos
\left(
\frac{{A}}{F}
\right) 
\right]
+
\ldots
\,,
\label{eq:S4}
\end{equation}
where
\begin{equation}
{A} (x) := 
\tilde{A^1}(x,\vec{0})\,,
\label{eq:phi}
\end{equation}
and
\begin{align}
V_4 &:= (2\pi L)^d V_{(4+d)} \,,
\label{eq:V4}\\
F &:=
f^c
(2\pi L)^{\frac{d}{2}}\,.
\label{eq:Fc}
\end{align}
For the sake of comparison,
in the (de)constructed model
we take all $d$ directions to be the same,
$f_I=f$, $N_I = N$ and $\gamma_I = \gamma$
($I = 1,2,\cdots, d$).

Comparing the 
$4$D potential 
with the original $(4+d)$D potential, 
we observe that:
1.~The overall height of the potential is 
multiplied by the factor $(2\pi L)^d$.
2.~The overall shape of the potential is elongated
in the field direction
by the factor $(2\pi L)^{\frac{d}{2}}$.

On the other hand, in the model with
(de)constructed extra dimensions,
by comparing the original potential \eqref{eq:AIset0}
with the final potential \eqref{eq:A1nset0}
we observe that:
1.~The overall height of the potential is 
multiplied by the factor $N^d$.
2.~The overall shape of the potential is elongated
in the field direction by the factor $N^{\frac{d}{2}}$.

Noticing that 
$L$ is proportional to $N$ when
$f$ and $g$ do not scale with $N$ (see \eqref{eq:L})
in the model with (de)constructed extra dimensions,
we observe that there is a parallel
between the continuous extra dimensions
and (de)constructed dimensions.
The correspondence between
them
are summarized in Table.~\ref{table:DeCont}.

However, in spite of the enhancement 
in the field range,
in the case of continuous extra dimensions
the field range cannot exceed 
$4$D Planck scale
if 
the field range in higher dimensional theory
is bounded by higher dimensional Planck scale:
\begin{align}
F 
= 
f^c
(2\pi L)^{\frac{d}{2}}
\lesssim&
\,
M_{(4+d)}^{\frac{d+2}{2}} (2\pi L)^{\frac{d}{2}} 
=
\left(
\frac{M_P^2}{(2\pi L)^{d}}
\right)^{\frac{1}{{d+2}}\cdot \frac{d+2}{2}}
(2 \pi L)^{\frac{d}{2}}
=M_P \,.
\label{eq:DRDC}
\end{align}

This is the place
where the difference between
the model with continuous extra dimensions
and that with (de)constructed extra dimensions comes in.
Since the
(de)constructed model is purely a $4$D QFT,
there is no notion of 
higher dimensional Planck scale.
Thus unlike $f^c$ in the model 
with continuous extra dimensions
which may be bounded from above by the 
$(4+d)$D 
(reduced) Planck scale $M_{(4+d)}$,
the symmetry breaking scale $f$ 
is bounded from the above only by 
the $4$D (reduced) Planck scale $M_P$.
This allows the effective field range 
to be trans-Planckian 
in models with (de)constructed extra dimensions
if we take $N$ sufficiently large.
Also note that
one cannot take the continuum limit $N \rightarrow \infty$
in (de)constructed model.
The gauge coupling $g$ should also be bounded 
from the above as $g \lesssim 1$,
otherwise
the perturbative description of the 
action \eqref{eq:decS5} is invalid.
Then from \eqref{eq:L},
$N \rightarrow \infty$ leads to 
the decompactification
$L \rightarrow \infty$,
which is not acceptable.

%

The period of the Wilson loop operator 
wrapping around $y^1$ direction
and
the period of the mass term of the charged field
is determined by the gauge transformation
\begin{equation}
\alpha(\vec{y})
=
\frac{y^1}{g_{(4+d)} L} \,,
\label{eq:calpha}
\end{equation}
which leads to the gauge equivalence
\begin{equation}
A^1(\vec{y}) \sim A^1(\vec{y}) + \frac{1}{g_{(4+d)} L} \,.
\label{eq:A1sim}
\end{equation}
From \eqref{eq:FText},
\eqref{eq:A1sim} leads to the 
gauge equivalence of the zero-mode
of $A^1(\vec{y})$:
\begin{equation}
\frac{1}{(2\pi L)^{\frac{d}{2}}}
A
\sim
\frac{1}{(2\pi L)^{\frac{d}{2}}}
A +
\frac{1}{g_{(4+d)} L} \,,
\label{eq:Asim}
\end{equation}
which can be rewritten as
\begin{equation}
A \sim A + 2\pi f_4 \,,
\label{eq:Asim2}
\end{equation}
where $f_4$ is given by
\begin{equation}
f_4 
=
\frac{(2\pi L)^{\frac{d}{2}-1}}{g_{(4+d)}}
=
\frac{1}{g_4 2\pi L} \,.
\label{eq:f4c}
\end{equation}
Again we observe 
that the $L$ dependence of $f_4$
in the model with continuous extra dimensions
is in parallel
with the $N$ dependence of $f_4$
in the (de)constructed model.

\begin{table}[H]
\begin{center}
{\renewcommand{\arraystretch}{2.0}
  \begin{tabular}{ | c | c | }
    \hline
   {(De)constructed} & {Continuous} \\ \hline 
	 $\displaystyle 2\pi L = Na = \frac{N}{gf}$ & $2\pi L$ \\[2mm] \hline
   $a^{-\frac{d}{2}} A^1_{(\vec{j},\vec{j}+\vec{e}_1)}$ & $A^1(\vec{y})$\\ \hline
	 $\displaystyle A = \frac{1}{N^{\frac{d}{2}}} \sum_{\vec{j}} A^1_{(\vec{j},\vec{j}+\vec{e}_1)} $ 
 & $\displaystyle A = \frac{1}{(2\pi L)^{\frac{d}{2}}}\int d^dy A^1(y)$\\ \hline
   $a^{\frac{d}{2}} g$ & $g_{(4+d)}$  \\ \hline
	 $\displaystyle g_4 = \frac{g}{N^{\frac{d}{2}}}$ 
 & $\displaystyle g_{4} = \frac{g_{(4+d)}}{(2\pi L)^{\frac{d}{2}}}$  \\[3mm] \hline
	 $a^{-\frac{d}{2}} f$ & $f^c$  \\ \hline
	 $\displaystyle f_4 = f N^{\frac{d}{2}-1} $ 
 & $\displaystyle f_4 = \frac{(2\pi L)^{\frac{d}{2}-1}}{g_{(4+d)}}$  \\ \hline
   $F = f N^{\frac{d}{2}}$ & $F = f^c (2\pi L)^{\frac{d}{2}}$  \\ \hline
	 $a^{-{d}} \tilde{V}_{1}(p_1=1)$ & $V_{(4+d)}$ \\ \hline
	 $\displaystyle V_4 = N^{\frac{d}{2}}\tilde{V}_{1}(p_1=1) 
	 $ 
 & $\displaystyle  V_{4} 
    = (2\pi L)^d V_{(4+d)} 
		$ \\ \hline
  \end{tabular}
	\caption{The correspondence between the 
	quantities in (de)constructed extra dimensions and 
	those in continuous extra dimensions.
	The correspondence is such that
  the quantities in the left column
	goes to the quantities in the right column
	in the formal continuum limit
	$N \rightarrow \infty$, $a \rightarrow 0$ with $N a = 2\pi L$ fixed. 
	Here, $a := 1/ g f$ 
	in terms of the original parameters in 4D QFT.}\label{table:DeCont}
	}
\end{center}
\end{table}

\section{An explicit inflation model and
comparison with the observations}\label{sec:Model}

In this section we construct 
a single field slow-roll inflation model
in which the field $A$,
the ``zero-mode'' of a component of 
the gauge field in (de)constructed extra dimensions, 
plays the role of the inflaton.
For simplicity,
we consider the case 
in which all $d$ directions in 
the (de)constructed extra dimensions
look
the same,
$f_I = f$, $N_I=N$ 
and $\gamma_I = \gamma$ $(I=1,2,\cdots,d)$.

We consider the case where
the Stueckelberg potential 
\eqref{eq:Vphi}
dominates
over the Wilson loop potential
\eqref{eq:WLpot}:
\begin{align}
V_{WL}(A) &\ll V_{S}(A) \,,
\label{eq:looptree1}\\
V_{WL}'(A) &\ll V_{S}'(A) \,.
\label{eq:looptree2}
\end{align}
\eqref{eq:looptree1} and \eqref{eq:looptree2} 
respectively give
\begin{align}
f 
&\lesssim 
2 \times 10^{-3} (4\pi)^\frac{N}{2} M_P\,,
\label{eq:looptree1b}\\
f 
&\lesssim 
1 \times 10^{-4} 
\frac{(4\pi)^{\frac{2N}{3}}}{N^{\frac{d}{6}}} M_P\,.
\label{eq:looptree2b}
\end{align}
Here, we have used
$c_{WL} \sim \Ord(1)$ and $\Lambda = 4\pi f$.
With a moderately small lattice size
$N\sim 6$ and above, 
\eqref{eq:looptree1b} and \eqref{eq:looptree2b}
do not give further constraint
once $f \ll M_P$ is assumed.

So far,
the Stueckelberg potential 
\eqref{eq:A1nset0} 
was an arbitrary symmetric function of $A$ in the field range
$- \pi F \leq A < \pi F$.
Now, as an example,
we assume that 
the inflation took place
in the region of the potential which is approximately linear:
\begin{equation}
    V_S(A) = \lambda M_{P}^3 | A-A_0 | \,,
    \label{eq:Vphi}
\end{equation}
where $A_0$ is a positive constant with
$A_0 \ll \pi F$.
Furthermore, without loss of generality,
we assume that the inflation took place 
in the region $A > A_0$.
We define $\phi = A-A_0$
and then the inflaton potential can be written 
in the field range of our interest as
\begin{equation}
V(\phi) = \lambda M_{P}^3 \phi \,.
\label{eq:VSphi}
\end{equation}
The linear 
inflaton potential is compatible
with the latest Planck 2018 results
\cite{Akrami:2018odb}.
From the inflaton potential \eqref{eq:VSphi},
we obtain the slow-roll parameters as
\begin{align}
\epsilon(\phi)  
&:= 
\frac{M_{P}^2}{2}
\left(
\frac{V^\prime}{V}
\right)^2
= 
\frac{M_{P}^2}{2\phi^2} \,,
\label{eq:eps}\\
\eta(\phi) 
&:= 
{M_P^2}
\frac{V''}{V} = 0 \,.
  \label{eq:eta}
\end{align}
The number of e-folds is given as
\begin{equation}
N(\phi) 
\simeq 
\int_{\phi_{end}}^{\phi} d\phi 
\frac{V}{M_P^2 V^\prime} 
= 
\frac{1}{M_{P}^{2}}
\left[
\frac{\phi^{2}}{2}
\right]_{\phi_{end}}^{\phi}\,,
\label{eq:Ne}
\end{equation}
where we define the inflaton 
field value $\phi_{end}$ 
at the end of the slow-roll inflation
by the condition 
$\epsilon (\phi_{end}) = 1$,
which gives
\begin{equation}
    \phi_{end}= \frac{M_{P}}{\sqrt{2}} \,.
\end{equation}
We choose the number of e-folds 
at the pivot scale $0.05$ Mpc$^{-1}$ as
\begin{equation}
    N(\phi_\ast) = 50 \,,
\end{equation}
where $\ast$ denotes the value at the pivot scale.
From \eqref{eq:Ne} we obtain
\begin{equation}
\phi_\ast \simeq 10 M_{P}\,.
\label{eq:phipvt}
\end{equation} 
The slow-roll parameters 
at the pivot scale are given as
\begin{align}
\epsilon(\phi_\ast) 
&\simeq 
5.0 \times 10^{-3} \,,
\label{eq:epspvt}\\
\eta(\phi_\ast) 
&= 
0 \,.
\label{eq:etapvt}
\end{align}
The scalar power spectrum is given as
\begin{equation}
P_s \simeq 
\frac{V(\phi_\ast)}{24 \pi^2 M_P^4 \epsilon(\phi_\ast)}
= 2.2 \times 10^{-9} \,,
\label{eq:Ps}
\end{equation}
where the value on the right hand side is 
the COBE normalization.
Substituting 
\eqref{eq:VSphi}, 
\eqref{eq:phipvt}, 
\eqref{eq:epspvt}
in \eqref{eq:Ps}
the value of the parameter $\lambda$ is fixed as
\begin{equation}
    \lambda \simeq 2.6 \times 10^{-10} \,.
\label{eq:lambda}
\end{equation}
The scalar spectral index $n_s$
is given as
\begin{equation}
n_s \simeq 1 - 6 \epsilon(\phi_\ast) + 2 \eta(\phi_\ast)
=
0.97\,.
\label{eq:SSI}
\end{equation} 
The tensor-to-scalar ratio 
$r_\ast$ at the pivot scale 
is given as
\begin{equation}
r_\ast \simeq 16 \epsilon(\phi_\ast)
= 
8.0 \times 10^{-2} \,.
\end{equation}
From the slow-roll approximation of 
the Friedmann equation,
\begin{equation}
    H(\phi)^{2} \simeq \frac{V(\phi)}{3M_{P}^2} \,,
\end{equation}
we obtain the Hubble scale when 
the pivot scale exited the horizon:
\begin{equation}
H(\phi_{\ast}) \simeq 7.2 \times 10^{13}\, \mathrm{GeV}\,.
\end{equation}
%

Now, 
in order to achieve the trans-Planckian inflaton excursion
\eqref{eq:phipvt},
the lattice size $N$ need to be large enough
so that
the field range $2\pi f N^{\frac{d}{2}}$
can accommodate the excursion.
This condition requires the lowest value of $N$ as
\begin{equation}
N 
\gg 
(3.9 \times 10^2 )^{\frac{2}{d}}
\times
\left(
\frac{f}{1.0\times 10^{16} \,\mathrm{GeV}}
\right)^{-\frac{2}{d}} \,.
\label{eq:Nlb}
\end{equation}
The constraint becomes weaker for larger $d$,
in which case a small value of $f$ may be accommodated.

In order for the 
inflaton EFT \eqref{eq:decS4} to be valid
during inflation,
the ``KK" energy scale must be above
the Hubble scale at the time of inflation. 
This condition gives an upper bound on the lattice size $N$:
\begin{equation}
N 
<
8.8 \times 10^2
\times
\left(
\frac{g}{1.0}
\right)
\left(
\frac{f}{1.0\times 10^{16} \, \mathrm{GeV}}
\right)\,.
\label{eq:Nub}
\end{equation}
When $d=1$ with $g \simeq 1$,
in order for the lower bound \eqref{eq:Nlb}
to be below the upper bound \eqref{eq:Nub},
$f$ needs to be as large as $\simeq 6 \times 10^{16}$ GeV,
which 
is getting close to the bound
$\Lambda = 4\pi f \ll M_P$.
The constraints are mild for $d \geq 2$.

The $A$ dependent mass \eqref{eq:Mchi2n}
of the charged field $\tilde{\chi}_{(n)}$
may have an interesting cosmological consequence,
if $m^2 - 2 \gamma f^2 \ll H^2$ during inflation:
Whenever the inflaton
$A$ crosses the value
$A = - 2 \pi f_4 n$ 
the field $\tilde{\chi}_{(n)}$ becomes almost massless,
leading to a burst of productions
of $\tilde{\chi}_{(n)}$ particles,
which may leave observable features
in the anisotropy of the CMB radiation
\cite{Chung:1999ve,%
Barnaby:2009mc,Barnaby:2009dd,%
Pearce:2017bdc}.
This is an extension of the mechanism
``\textit{excursions through KK modes}''
found in \cite{Furuuchi:2015foh}
to the case where the KK modes 
are those of the (de)constructed extra dimensions.

Near $A = - 2\pi f_4n$,
the 
mass term of $\tilde{\chi}_{(n)}$
can be approximated as
\begin{equation}
\tilde{\chi}_{(n)}^\dagger
M^2_{\chi(n)}(A)
\tilde{\chi}_{(n)}
\simeq
\tilde{\chi}_{(n)}^\dagger
\gamma
\left(
\frac{A + 2\pi f_4 n}{N^{\frac{d}{2}}}
\right)^2
\tilde{\chi}_{(n)}\,.
\label{eq:chimassppp}
\end{equation}
Here, we have assumed $m^2 - 2 \gamma f^2 \ll H^2$.
\eqref{eq:chimassppp}
is the same interaction
studied in
\cite{Barnaby:2009mc,Barnaby:2009dd,Pearce:2017bdc}.
From the latest analytic result of
\cite{Pearce:2017bdc},
the contribution of the rapid particle production
due to the interaction \eqref{eq:chimassppp}
to the power spectrum $\delta P_s$ is given by
\begin{equation}
\delta:=
\frac{\delta P_s}{P_s}
\simeq
2\times 300 
\left(
\frac{\sqrt{\gamma}}{N^{\frac{d}{2}}}
\right)^{7/2} \,,
\label{eq:delta}
\end{equation}
where the factor $2$
in the right hand side
came from the fact that
the complex field $\chi_{(n)}$ has
two real degrees of freedom.
Below,
we restrict ourselves to 
the case
close to the 
universality restoration point
$
\gamma \simeq g^2 
$
(see the explanation below
\eqref{eq:Mchi2n})
in order to make a direct comparison with the model 
with continuous extra dimension \cite{Furuuchi:2015foh}.
As a crude upper bound
on $\delta$,
we assume that the 
contribution of the particle production
to the power spectrum
does not exceed that of the inflaton.\footnote{%
When the distances between the peaks of 
the primordial features in the power spectrum are small,
i.e. $\Delta_i \ll 1$ where
$\Delta_i$ is defined in \eqref{eq:DeltaN},
observations will not be able to resolve
each peak \cite{Barnaby:2009dd}.
In such a case it may become harder 
to distinguish 
the primordial features in the power spectrum from
the almost scale invariant power spectrum.}
As a rough criterion
for the detectability of
the primordial feature 
in near future,
we require that
the amplitude of the feature
to be more than a percent.  
These requirements,
\begin{equation}
0.01 < \delta < 1 \,,
\label{eq:deltaub}
\end{equation}
give
\begin{equation}
(6.2\, g)^{\frac{2}{d}}
<
N
<
(2.3 \times 10\, g)^{\frac{2}{d}}\,.
\label{eq:Nbddelta}
\end{equation}
It is natural to 
have the value of the gauge coupling 
in the range
$0.1 \lesssim g \lesssim 1$.
Then, the upper bound on $N$ in \eqref{eq:Nbddelta}
is quite tight when $d \geq 2$.
However, note that
the upper bound of 
\eqref{eq:Nbddelta}
came from the condition
that the primordial feature is 
within the reach of near future detection.
Thus when $N$ is smaller than this bound, 
the 
primordial feature may not be detectable
in the near future,
but the model is still valid
as a single field slow-roll inflation model.

The number of e-folds from
the $i$-th peak to the $(i+1)$-th peak
is given by
\begin{align}
\Delta_i
&:=
\ln \left( \frac{k_{i+1}}{k_{i}}\right)
\nn\\
&=
N(\phi(k_{i})) - N(\phi(k_{i+1})) 
\nn\\
&= 
N(\phi(k_{i})) - N(\phi(k_{i}) - 2\pi f_4) 
\nn\\
&\simeq
\frac{dN}{d\phi}
(\phi(k_i))
{2\pi f_4}
\nn\\
&=
\frac{\phi(k_i)}{M_P^2} {2\pi f_4}\,.
\label{eq:DeltaN}
\end{align}
Remembering \eqref{eq:f4decHD},
\eqref{eq:DeltaN} can be rewritten as
\begin{equation}
\Delta_i
= 
2.6 \times 10^{-1} \times
\left(
\frac{\phi(k_i)}{10 M_P}
\right)
\left(
\frac{f}{1.0 \times 10^{16} \, \mathrm{GeV}}
\right)
N^{\frac{d}{2}-1} \,.
\label{eq:DeltaN2}
\end{equation}
Notice that
for $d \geq 3$
a larger lattice size $N$ leads to
larger intervals between 
the peaks of the primordial features.

\section{Summary and 
discussions}\label{sec:Discussions}

In this article,
we constructed 
a massive gauge field theory
with (de)constructed extra dimensions.
One of our main results was
that
the effective field range of 
the ``zero-mode''
of a component
of the massive gauge field 
in 
the (de)constructed extra dimensions
was enhanced by a factor
$N^{\frac{d}{2}}$,
where $d$ was the number of the
(de)constructed extra dimensions and 
$N$ was the number of the lattice points
in each (de)constructed direction.
We applied 
this mechanism of field range enhancement
in a large field inflation model.
We obtained constraints on the model parameters
from the latest CMB observations.
We also extended 
the rapid particle production mechanism
``\textit{excursion through KK modes}''
to the case with ``KK'' modes of the
(de)constructed extra dimensions.
The cosmological consequences of this mechanism
were also studied and compared with the CMB data.

In this article
we focused on the case 
in which 
the Stueckelberg potential
dominates over
the Wilson loop potential, 
\eqref{eq:looptree1} and \eqref{eq:looptree2}.
However, it should be noted that
the enhancement of the period
also occurs 
in the Wilson loop potential
for $d \geq 3$, 
as shown in \eqref{eq:f4decHD}.
%
%
Therefore, for $d \geq 3$
we can have a large field inflation model
in which the Stueckelberg potential
is sub-dominant or even absent.
This model is a 
(de)constructed version of 
the original extra-natural inflation.

We observed in Sec.~\ref{subsec:cont}
that in a model with
continuous extra dimensions,
the field range cannot exceed the
4D Planck scale $M_P$
if the original field range
in the higher dimensional theory
is below the higher dimensional Planck scale.
The purely $4$D nature
of the (de)construction
circumvented this constraint.
Those familiar with string theory might worry
that
if our (de)constructed model is to be realized
in string theory,
the lattice of the (de)constructed space
may be embedded in real space-time
in which closed strings propagate.
Then the circumferences of the periodic lattice
may coincide with the circumferences of 
the real continuous extra dimensions,
and the obstacle for achieving 
an effective trans-Planckian field range
may reappear in the (de)constructed model.
While this is a valid concern,
we feel 
it is still too early to make a conclusion
on this issue,
since our current understanding of 
string vacua is still very limited.
For example, it is not clear whether 
(de)constructed space always need to be
embedded in real space-time in string theory.
We leave this interesting issue
to future investigations.

Dimensional (de)construction is not the only
way to have purely 
$4$D QFT description
of extra dimensions.
For example, a gauge-Higgs unification model
from spontaneously created fuzzy extra-dimensions
was proposed in \cite{Furuuchi:2011px}.
It will also be interesting to explore
inflation models based on fuzzy extra dimensions.

\vskip8mm
\begin{center}
\textit{Acknowledgments}
\end{center}
\vskip-2mm
This work is supported in part by 
the Science and Engineering Research Board,
Department of Science and Technology, Government of India
under the project file number EMR/2015/002471.
The work of Suvedha Suresh Naik
is supported by
Dr.~T.M.A.~Pai Ph.~D. scholarship program of 
Manipal Academy of Higher Education.
Manipal Centre for
Natural Sciences, Centre of Excellence,
Manipal Academy of Higher Education 
is acknowledged for facilities and support.

\appendix
\section{Discrete Fourier Transform}\label{App:DFT}

Let us consider a cyclically ordered $N$ points
labeled by $j$ 
($j = 0, 1, \cdots , N-1$ $(\mathrm{mod}\,\, N)$).
Consider 
a variable $f_j$ which has a value on each point.
We use the following convention for
the discrete Fourier expansion of the variable
$f_j$:
\begin{align}
f_{j} 
&= 
\frac{1}{\sqrt{N}}
\sum_{n=-\frac{N-1}{2}}^{\frac{N-1}{2}}
\tilde{f}_n\, e^{i\frac{2\pi  n j}{N}} 
\qquad (N: \mbox{odd})\,.
\label{eq:fnodd}
\\
f_{j} 
&= 
\frac{1}{\sqrt{N}}
\sum_{n=-\frac{N}{2}-1}^{\frac{N}{2}-1}
\tilde{f}_n\, e^{i\frac{2\pi  n j}{N}}
+
\frac{1}{\sqrt{N}}
\tilde{f}_{\frac{N}{2}}
(-)^j  
\qquad (N: \mbox{even})\,.
\label{eq:fneven}
\end{align}
Our convention is convenient
since when applied in (de)construction
each ``KK'' mode is canonically normalized.

When $f_j$ is a real variable,
$\tilde{f}_{-n}^\ast = \tilde{f}_n$.
The
orthogonality 
of the exponential function:
\begin{equation}
\sum_{j=0}^{N-1}
\left(
e^{i\frac{2\pi  n_1 j}{N}} 
\right)^\ast
e^{i\frac{2\pi  n_2 j}{N}}
= N \delta_{n_1 n_2}\, ,
\label{eq:DFTo}
\end{equation}
leads to the following formula for the
discrete Fourier coefficient:
\begin{equation}
\tilde{f}_n
=
\frac{1}{\sqrt{N}}
\sum_{j=0}^{N-1}
f_j e^{-i\frac{2\pi  n j}{N}}\,.
\label{eq:invDFT}
\end{equation}

We would like to have a formula for
the discrete counterpart of 
the dimensional reduction.
For this purpose,
let us first consider
\begin{align}
\sum_{j=0}^{N-1} 
(f_j)^m
&=
\sum_{j=0}^{N-1} 
\left(
\frac{1}{\sqrt{N}}
\sum_{n}
\tilde{f}_n e^{i\frac{2\pi  n j}{N}} 
\right)^m
\nn\\
&=
\frac{1}{N^{\frac{m}{2}}}
\sum_{j=0}^{N-1} 
\sum_{n}
\cdots
\sum_{n_m}
\tilde{f}_{n_1}
\cdots
\tilde{f}_{n_m}
\exp 
\left[ i \sum_{\ell=1}^m \frac{2\pi  n_\ell j}{N} \right]
\nn\\
&=
\frac{N}{N^{\frac{m}{2}}}
\sum_{n_1}
\cdots
\sum_{n_m}
\tilde{f}_{n_1}
\cdots
\tilde{f}_{n_m}
\delta_{n_1+n_2+\cdots n_m =0 \,\, \mathrm{mod}\,\, N}
\,.
\label{eq:fm}
\end{align}
In \eqref{eq:fm} the sum over
$n_\ell$ $(\ell = 1,2,\cdots,m)$
is taken as in \eqref{eq:fnodd} or 
\eqref{eq:fneven}
depending on whether $N$ is odd or even.

In the discrete version of the dimensional reduction,
we set
all the discrete Fourier coefficients except for the 
``zero-mode'' to zero:
$\tilde{f}_k =0$ for $k\ne 0$ 
in \eqref{eq:fnodd} or \eqref{eq:fneven}.
In this case,
\eqref{eq:fm}
becomes
\begin{equation}
\sum_{j=0}^{N-1} 
(f_j)^m
\Biggl|_{\tilde{f}_k =0 \,\,\mathrm{except}\,\, k=0}
=
N
\left(
\frac{\tilde{f}_{0}}{\sqrt{N}}
\right)^m \,.
\label{eq:DDR}
\end{equation}
Now suppose that
a function $V(x)$ 
has a Taylor series expansion around $x=0$:
\begin{equation}
V(x) = \sum_{m=0}^\infty \frac{V^{(m)}(0)}{m!} x^m \,,
\label{eq:FTE}
\end{equation}
where $V^{(m)}(0)$ denotes the $m$-th derivative
of the function $V(x)$ at $x=0$.
Consider a
field theory on the discrete points
with a potential of the form
$V(f_j)$.
Then from \eqref{eq:DDR} 
the discrete dimensional reduction
of this potential is given as
\begin{equation}
\sum_{j=0}^{N-1}
V(f_j)\Biggl|_{\tilde{f}_k =0 \,\,\mathrm{except}\,\, k=0}
=
N
\sum_{m=0}^\infty \frac{V^{(m)}(0)}{m!}
\left(
\frac{\tilde{f}_{0}}{\sqrt{N}}
\right)^m \,.
\label{eq:FTEf}
\end{equation}

\bibliography{N_Dec_Ref}
\bibliographystyle{utphys}
\end{document}